\documentstyle[12pt]{article}
\input{psfig}
\begin{document}
\title{\bf Transfer of coherence from atoms to mixed field states in a 
two-photon lossless micromaser}
\author{A.F. Gomes\thanks{Permanent address: 
Funda\c c\~ao Educacional de Barretos, C.P. 16, 14783-226, Barretos, SP, Brazil;
E-mail: afgomes@ifi.unicamp.br}, 
J.A. Roversi\thanks{E-mail: roversi@ifi.unicamp.br}, and A. Vidiella-Barranco
\thanks{Phone: +55 19 788 5442; Fax:+55 19 788 5427; E-mail: vidiella@ifi.unicamp.br}\\
{\it Instituto de F\'\i sica `Gleb Wataghin'},\\
{\it Universidade Estadual de Campinas},\\
{\it 13083-970   Campinas  SP  Brazil}}
\date{\today}
\maketitle
\begin{abstract}
We propose a two-photon micromaser-based scheme for the generation of a nonclassical 
state from a mixed state. We conclude that a faster, as well as a higher 
degree of field purity is achieved in comparison to one-photon processes. We 
investigate the statistical properties of the resulting field states, for 
initial thermal and (phase-diffused) coherent states. Quasiprobabilities are 
employed to characterize the state of the generated fields.
\end{abstract}

Pacs: 42.50.-p, 03.65.Bz, 42.50.Dv

\newpage

\section{Introduction}

The generation of pure states of the electromagnetic field is an important issue 
in quantum optics. Nonclassical states such as squeezed \cite{slus} and Schr\"odinger 
cat states \cite{brun} have been already generated, and several schemes for the 
generation of arbitrary quantum states, named quantum state engineering, 
have been proposed throughout. Normally those methods have the vacuum 
state $|0\rangle$ as a starting point for the field. The energy necessary to build 
up a given state may be supplied by atoms \cite{schl}, by coherent plus
one-photon fields \cite{wels}, as well as classical pumps \cite{eber,vidi}. In any 
case, photons are coherently added to an \underline{already pure state} (vacuum). It 
would be therefore
interesting to verify how quantum state generation could be achieved if we depart from
less favourable initial conditions, i.e., with initial mixed states, instead. This problem
has already been addressed in Ref. \cite{moya}, where it is described a scheme of 
purification of
thermal fields by means of one photon-transitions. The central point of the method is 
the progressive transfer of coherence from atoms to the cavity field. Atoms are prepared
in superpositions of circular Rydberg states (while crossing a Ramsey zone apparatus), 
and successively injected into a high-Q cavity, where the field, resonant with the 
atomic transition, is built up. The transfer of coherence from conveniently prepared 
atoms to cavity fields consists in an important mechanism for the investigation of quantum 
aspects of light and matter. It also leads to interesting effects, such as the atomic 
population trapping \cite{zahe,jona}, for instance. Besides, it is comparatively 
easier to prepare coherent superpositions of atomic states.

In this paper we are going to be concerned with the purification of mixed states in a
two-photon micromaser. In our scheme, conveniently prepared three-level atoms undergoing 
two-photon transitions are injected into a cavity. Obviously, we expect such a scheme 
to lead only to certain types of pure states, because of the effective two photon 
interaction.
Nevertheless, because for the same reason, as we are going to show, the cavity field 
purification may be attained faster than in a one-photon scheme. Moreover, in a
two-photon micromaser, a considerably higher degree of purification is achieved.

This paper is organized as follows; in section 2 we present our scheme of field 
generation in section 3 we discuss the statistical properties of the produced fields; 
in section 4 we show the evolution of the field according to the phase space 
representations; in section 5 we summarize our conclusions.

\section{General scheme}

We consider three-level atoms in a ladder-type configuration. The intermediate level may
be adiabatically eliminated, resulting the following effective Hamiltonian \cite{simon}
\begin{equation}
\hat H=\hbar \omega \hat a^+ \hat a + \frac{1} {2} \hbar (\omega_0+\chi \hat a^+ 
\hat a) \sigma_3 + \hbar \lambda (\hat a^{+2} \sigma_- + \hat a^2 \sigma_+),
\end{equation}
where $\lambda$ is the atom-field coupling constant; $\omega$ is the field frequency; 
$\omega_0$ the atomic transition frequency; $\sigma_3, \sigma_- ,\sigma_+$ are the atomic 
operators; $\hat a^+$, $\hat a$ the field operators, and the detuning $\Delta=\omega_0-
2\omega$. The Stark shift coefficient is $\chi$.
The evolution operator relative to the Hamiltonian above is straightforwardly obtained
\cite{sten}, resulting
\begin{equation}
\hat{U}(t)=
\left[
\begin{array}{cc}\hat{\alpha}_n(\gamma)& \hat{\beta}_n(\gamma)\\
\hat{\beta}_n(\epsilon) & \hat{\alpha}_n^{\dagger}(\epsilon)
\end{array}
\right],
\end{equation}
where

\begin{equation}
\hat{\alpha}_n(\gamma) = \cos(\hat \gamma_n \lambda t)+i \frac{\sin(\hat \gamma_n 
\lambda t)} {\hat \gamma_n} \left(\frac{\frac{\Delta}{2} + \chi \hat n}
{\lambda}\right)
\end{equation}

\begin{equation}
\hat{\beta}_n(\epsilon) = i \frac{\sin (\hat \epsilon_n \lambda t)} {\hat \epsilon_n}
\end{equation}
with
\begin{equation}
\hat \gamma_n^2=\left(\frac{\frac{\Delta}{2} + \chi \hat n}{\lambda}\right)^2
+(\hat n+1)(\hat n+2),
\end{equation}
and 
\begin{equation}
\hat \epsilon_n^2=\left(\frac{\frac{\Delta}{2} + \chi \hat n}{\lambda}\right)^2
+\hat n (\hat n-1).
\end{equation}
The time evolution of the total (atom-field) density operator $\hat{\rho}^{af}$ will be
\begin{equation}
\hat{\rho}^{af}(t)=\hat{U}(t)\hat{\rho}^{af}(0)\hat{U}^\dagger(t).
\end{equation} 
The successively injected atoms are prepared, before entering the cavity, in the 
following superposition of upper and lower levels
\begin{equation}
|\psi\rangle=a|g\rangle+be^{i\phi}|e\rangle,
\end{equation}
being $a$ and $b$ real (nonzero) numbers and $\phi$ a relative phase.
Therefore the atom will be, at $t=0$, in the pure state $\hat{\rho}^{a}(0)=|\psi\rangle
\langle\psi |$, while the field is in a mixed state $\hat{\rho}^{f}(0)$. 
We assume that the total initial state may be factorized, or 
$\hat{\rho}^{af}(0)=\hat{\rho}^{a}(0)\otimes\hat{\rho}^{f}(0)$.
As usual, the field state is readily obtained by tracing over the atomic variables, or
$\hat{\rho}^{f}(t)=Tr_a\left[\hat{\rho}^{af}(t)\right]$. 

After injecting $N$ atoms, the matrix elements of the field density operator in the number
state basis $\langle n|\hat{\rho}^{f}|n'\rangle=\rho_N ^f(n,n')$ will obey the
following (micromaser) recurrence formula 
\begin{eqnarray}
\rho_N ^f(n,n')&=&[b^2 \alpha_n(\gamma)\alpha_{n'}^{\dagger}(\gamma) 
+a^2\alpha_n(\epsilon)
\alpha_{n'}^{\dagger}(\epsilon)]\rho_{N-1} ^f(n,n')\nonumber\\
&+&a^2\beta_n(\gamma)\beta_{n'}(\gamma)\sqrt{(n+2)(n+1)(n'+2)(n'+1)} 
\rho_{N-1} ^f(n+2,n'+2)\nonumber\\
&+&b^2 \beta_n(\epsilon)\beta_{n'}(\epsilon)\sqrt{n(n-1)n'(n'-1)}
\rho_{N-1} ^f(n-2,n'-2)\nonumber\\
&+&iab e^{i\phi}\alpha_n(\gamma)\beta_{n'}(\gamma)\sqrt{(n'+2)(n'+1)}
\rho_{N-1} ^f(n,n'+2)\nonumber\\
&+&iab e^{-i\phi}\alpha_n(\epsilon)\beta_n(\epsilon)\sqrt{n'(n'+1)}
\rho_{N-1} ^f(n,n'-2)\nonumber\\
&-&iab e^{-i\phi}\beta_n(\gamma)\alpha_{n'}^{\dagger}(\gamma)\sqrt{(n+2)(n+1)}
\rho_{N-1} ^f(n+2,n')\nonumber\\
&-&iab e^{i\phi}\beta_n(\epsilon)\alpha_{n'}^{\dagger}(\epsilon)
\sqrt{n(n-1)}\rho_{N-1} ^f(n-2,n').\label{heq}
\end{eqnarray}
From this matrix elements which represent the state of the field, we are able 
to determine under which conditions we may attain a reasonable purification of the field,
namely, departing from a mixed state. We note that because $a$ and $b$ are both nonzero, 
off-diagonal elements (number state basis) will be generated, characterizing the 
transfer of coherence. The time-dependent field purity parameter $\zeta$, defined 
as
\begin{equation}
\zeta=1-Tr\left[(\rho^f){}^2\right]=1-\sum_{n,n'}|\rho^f_N(n,n')|^2,
\end{equation}
will provide us the necessary guidance in order to purify the initial field.
If $\zeta=0$, this means that $\hat{\rho}$ represents a pure state. Therefore, we   
now seek optimum conditions for field purification, which corresponds to values of 
$\zeta$ as close to zero as possible. 
It would be also interesting to build up the field, i.e., to 
increase the mean energy of the field while the atoms cross the cavity. 
Here we are going to consider two different initial fields, the thermal state 
(mean photon number $\overline{n}$)
\begin{equation}
\hat \rho^f_{th}(0)=\sum_{n=0}^\infty\frac{\overline{n}^{n}}{(\overline{n}+1)^{n+1}} 
|n\rangle \langle n|,\label{ther}
\end{equation}
and the mixed (phase diffused) coherent state
\begin{equation}
\hat \rho^f_{co}(0)=\sum_{n=0}^\infty \frac{|\alpha|^{2n} e^{-|\alpha|^2}}{n!} 
|n\rangle \langle n|.\label{coher}
\end{equation}
The overall features of the process are similar in either case, but important 
differences concerning the statistical properties of the field arise during the 
evolution.

\section{Results}

First we inspect the time-evolution of the field purity for one atom,  
choosing an interaction time that leaves the field purer as the atom exits the
cavity. Thus the next atom entering the cavity will interact with a `less mixed 
field'. In figure 1 we have plots of $\zeta$ as a function of time, for an initial 
thermal field with $\overline{n}=10$, after having passed 1, 20 and 100 atoms.
We note that for several ranges of times, the field is purer than the 
initial (mixed) state. In order to choose an optimum interaction time, we examined
the steady state, when $N$ atoms had already crossed the cavity, and imposing the 
condition that the final field state should be
considerably purer than the initial state. A second constraint is that the mean
photon number inside the cavity should either increase or remain constant. For
simplicity we have chosen the same interaction time for all atoms. This was the
main guidance for choosing that time. After calculating the field evolution for a 
range of times, we found that the optimum interaction time turns out to be 
$T\approx 12.2/\lambda$. The atoms are assumed to be prepared in a equally weighed 
state ($a=b=1/\sqrt{2}$), $\Delta=\chi=\lambda$, and $\phi=0$.  

In figure 2 we have a plot of the field purity ($\zeta$) of the final state, as
a function of the (scaled) interaction time $\lambda T$. We note that the field 
purity is maximum ($\zeta\approx 0$) for certain times.
However, this corresponds to the case in which photons are subtracted from the
cavity field, leading to either the vacuum state or the one-photon state. This is
seen in figure 3, where we have the mean photon number of the cavity field as a
function of the interaction times. The dashed line indicates the interaction time
$T\approx 12.2/\lambda$ we have chosen for our scheme. 

After fixing the interaction time, we may now analyze how the field changes as the
atoms successively enter the cavity.
In figure 4 we have a plot of $\zeta$ as a function of the number of atoms, for
both thermal and mixed coherent fields. We note that in either case the increase of 
purity (decrease of $\zeta$) is substantial even for $N=100$ atoms, and 
saturation ($\zeta\approx 0.53$) already exists around $N=200$ atoms, i.e., 
further injection does not improve the situation. We note that a larger degree of
purity is achieved faster for a phase diffused coherent state than for the 
thermal state, as we see in figure 4. 

This may be understood if we compare both 
initial photon number distributions with the distribution of the steady state field. 
Although there are not non-diagonal elements in the density matrix (number state 
basis) for any of the initial fields, the phase-diffused coherent
state has a more symmetrical (Poissonian distribution) than the thermal state
(geometrical distribution). In figure 5 we have a plot of the photon number 
distribution of the steady state for an initial thermal field. We note that
it has the overall shape similar to a Poissonian, the distribution of the phase
diffused coherent state, apart from the strong oscillations. It is therefore 
reasonable to expect the final state to be achieved more easily in that case, 
rather than for a thermal state, result which is apparent in the purity curve.
In figure 6 it is shown the mean photon number $\langle\hat{n}\rangle\equiv
\sum \rho^f_N(n,n)n$ in the cavity as a function of the number of atoms. 
We note the increase of energy from the 
initial $\langle n \rangle=10$ up to $\langle n \rangle\approx 32$ photons, and 
occurs
in a similar way for both thermal and mixed coherent initial fields.
 
Another interesting aspect is that the generated fields not only become purer than
the initial ones, but are also nonclassical in the sense that they may display 
sub-Poissonian statistics and/or
anti-bunching. This fact is represented in figure 7, where Mandel's Q parameter,
defined as $Q=\Delta\hat{n}^2/\langle n \rangle-1$ is 
plotted as a function of the number of atoms $N$. We note that for both an initial 
thermal field (figure 7 (a)), and for a mixed coherent state ((figure 7 (b)), 
the field becomes sub-Poissonian after around 100 atoms have crossed the cavity. 
However, there is an important difference between the two
cases for a not so large number of atoms. For an initial thermal state the field 
becomes monotically less super-Poissonian, while for an initial mixed coherent state 
(it is Poissonian at $t=0$) it starts becoming super-Poissonian, i.e., the parameter
Q grows until it reaches a maximum value (after passing around 30 atoms). Then
it starts decreasing, turning sub-Poissonian even faster than the thermal state case, 
as it is shown in figure 7. In both cases the field becomes anti-bunched after 
injecting more than 100 atoms. The generated field also displays strong oscillations 
in its photon number distribution (see figure 5) which characterizes Schr\"odinger 
cat-like states \cite{vidi1}.

We could then seek for more details about the generated state. 
Because we have obtained the density operator in the number state basis, 
it is convenient to switch to other representations, such as the phase-space 
quasiprobabilities.

\section{Phase-space approach}

Quasiprobability distributions have become important tools not only for quantum
state characterization, but have also been playing an active role in quantum state
reconstruction \cite{ulf}. Here we are going to be concerned with the characterization
of the field produced in the cavity. Specially useful in this case is the series
representation of quasiprobabilities \cite{moya1}
\begin{equation}
P(\beta;s)=\frac{2}{\pi}\sum_{k=0}^\infty(-1)^k\frac{(1+s)^k}{(1-s)^{k+1}}
\langle\beta,k|\hat{\rho}^f_N|\beta,k\rangle.
\end{equation}
We may represent the field density operator as $\hat{\rho}^f_N=\sum_{n,n'}
\rho^f_N(n,n')|n\rangle\langle n'|$, where the matrix elements $\rho^f_N(n,n')$ are
the ones in equation \ref{heq}. For $s=1$ we have the Q function, and $s=0$ we obtain the
Wigner function. In figure 8 we have a plot of the contours of the Q function after 
passing $N=200$ atoms through the cavity, for a field initially prepared in a thermal
state. We note that four peaks are formed around 
the origin, indicating that a certain symmetry of the quasiprobabilities around the origin 
is preserved during the generation process. We would also like to mention that with 
a convenient choice of parameters, in such a
way that photons are subtracted, we may also reach a pure field, e.g., a one-photon 
state. Nevertheless this case is not as interesting as if we are able to purify the
field and build it up at the same time, as we have shown above. 

The quasiprobabilities approach strongly suggests that the steady-state field 
is constituted by a superposition of four deformed Gaussian-like distributions, 
which resemble squeezed state's ones. We may therefore conjecture that the resulting 
state is in fact some kind of superposition of four squeezed coherent states. 
This is also supported by the fact that the photon number distribution displays 
strong oscillations, characteristic of those kind of superpostions \cite{vidi2}. 
Because our method is based on two-photon interactions, which are necessary for 
squeezed state generation, we could have expected the generation of fields somehow
related to squeezed states. We would like to remark that our aim was to find a way of 
allowing the generation of the purest possible state, even departing from 
highly mixed states, instead of establishing a more general quantum state 
engineering scheme.

\section{Conclusions}

We have investigated the process of generation of a nonclassical state departing 
from a thermal state by means of a two-photon micromaser. 
A reasonable degree of field purity is quickly achieved, and
the generated field presents nonclassical properties such as sub-Poissonian character.
Therefore the purification procedure also leads to the generation of a nonclassical 
state. We should remark that after passing about only $N=100$ atoms in the cavity, we
were able to obtain a degree of purity ($\zeta\approx 0.53$) higher than in 
one-photon transitions ($\zeta_{op}\approx 0.65$) \cite{moya}. Of course
the more intense the initial field is, the more difficult will be to
purify it as time goes on. Here we have attained a good degree of field purity
even having started with a relatively noisy field, containing around $\overline{n}=10$ 
thermal photons. We have studied the case in which the mean number of photons of 
the field is increased. 
We have so far identified two types of states; the four-peaked distribution shown in
figure 8, and the trivial case (photons are subtracted), which leads to either to the
vacuum state or the one-photon state. In both cases the states have a representation in
phase space symmetric in relation to the origin, similarly to the initial states. 
We have neglected losses as an approximation, and of course we expect decay to 
compete against the generation process. Nevertheless we would like to make a 
couple of remarks; the total estimated time for the experiment is around 
five times longer than the energy decay time in a state-of-the-art high $Q$ cavity. 
In the beginning of the process, one-photon losses are not expected to be of much 
importance, given the initial states, which will not have their statistics 
substantially changed by decay. However, as times goes on, decay will surely not 
favour the process. Nevertheless the investigation of ideal situations surely 
enlightens the discussions on the question of quantum state generation.

\newpage

\noindent {\bf\LARGE Acknowledgements}

\vspace{0.3cm}

Two of us, AFG and JAR, would like to acknowledge financial support from 
Conselho Nacional de Desenvolvimento Cient\'\i fico e Tecnol\'ogico (CNPq), Brazil.

\newpage

\begin{figure}[hp]
\vspace{1cm}
\centerline{\hspace{1.0cm}\psfig{figure=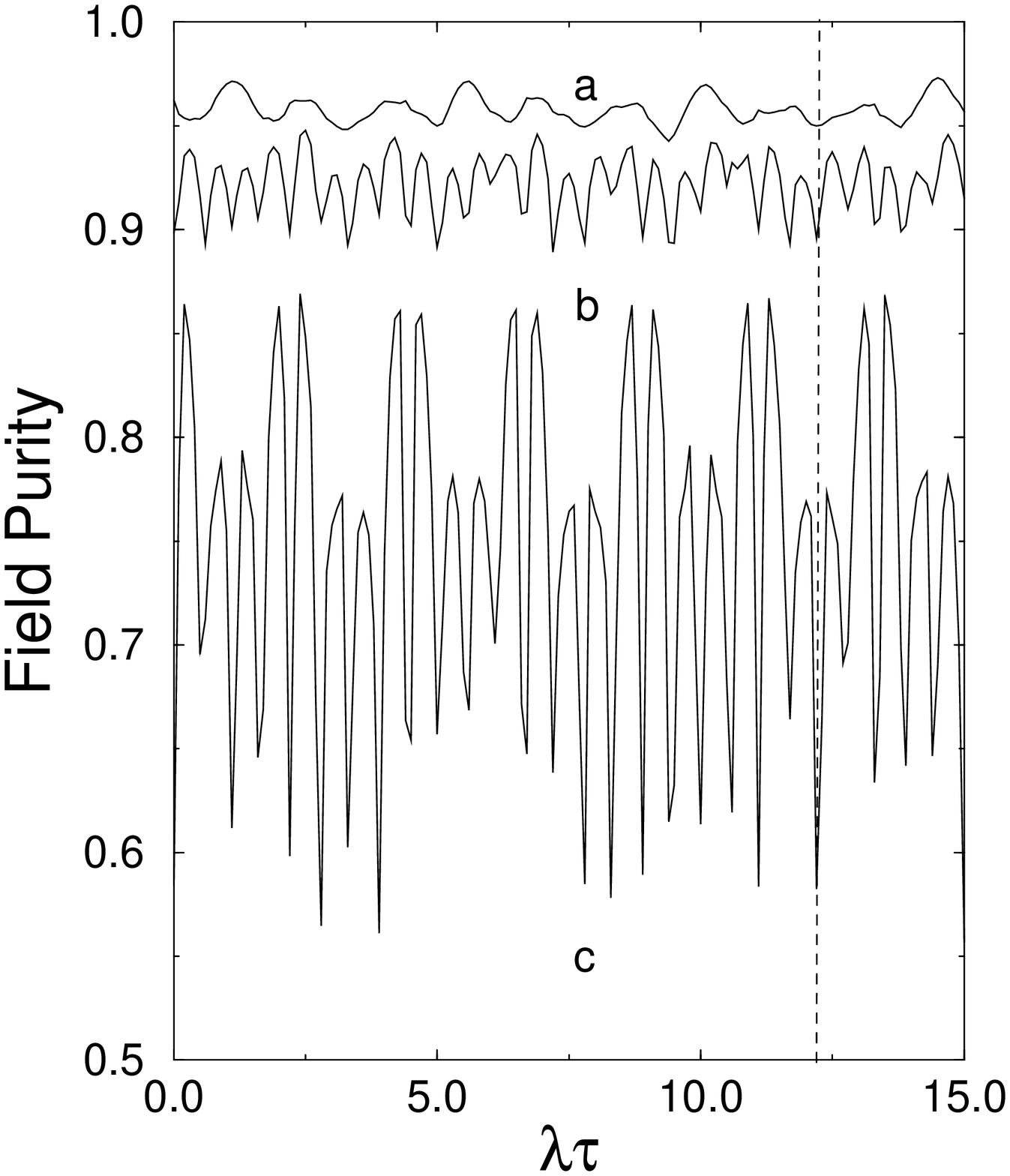,height=12cm,width=8cm}}
\vspace{2cm}
\caption{Purity of the field, as a function of time, for an initial thermal 
state having $\overline{n}=10$ and with $a^2=b^2=1/2$, 
$\Delta=\chi=\lambda$, and $\phi=0$., after having passed (a) N= 1 atom, 
(b) N=20 atoms and (c) N= 100 atoms. Note that in every case 
$t=(N-1)12.2/\lambda+\tau$.
The dashed line indicates the optimum interaction time $T=12.2/\lambda$.}
\end{figure}

\newpage

\begin{figure}[hp]
\vspace{1cm}
\centerline{\hspace{1.0cm}\psfig{figure=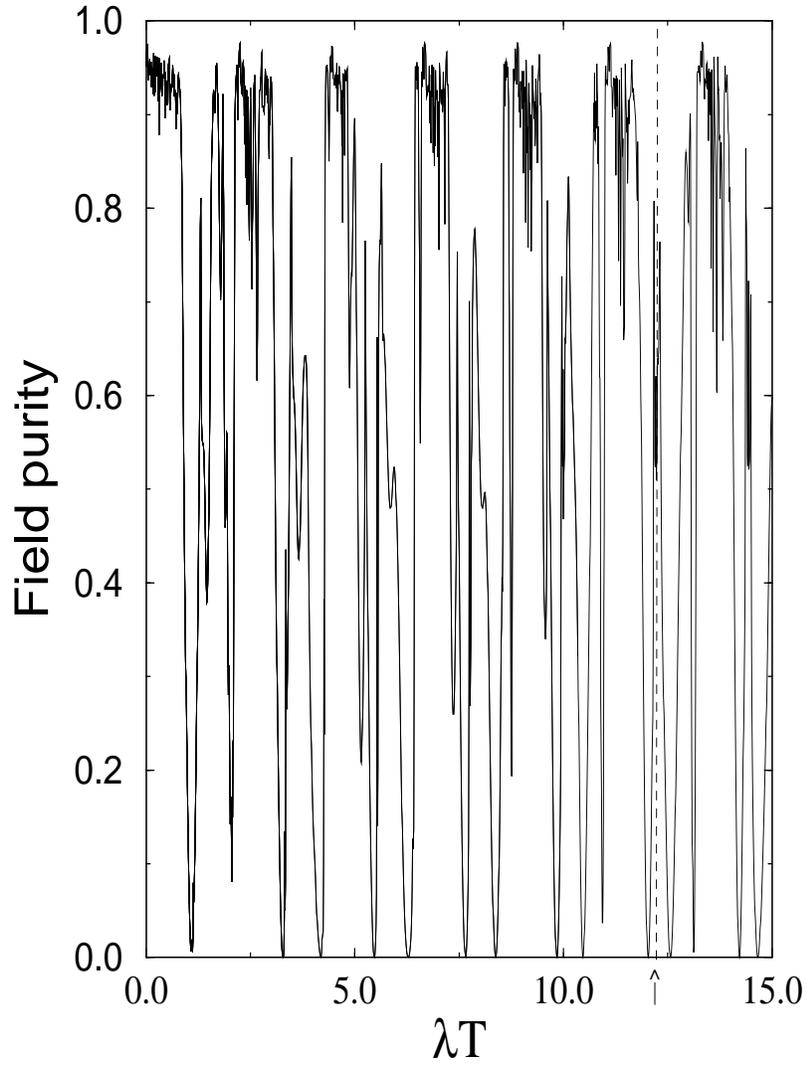,height=12cm,width=8cm}}
\vspace{2cm}
\caption{Purity of the steady state field, as a function of the interaction times.
The dashed line indicates the optimum interaction time $T=12.2/\lambda$.
The same parameters as in figure 1.}
\end{figure}

\newpage

\begin{figure}[hp]
\vspace{1cm}
\centerline{\hspace{1.0cm}\psfig{figure=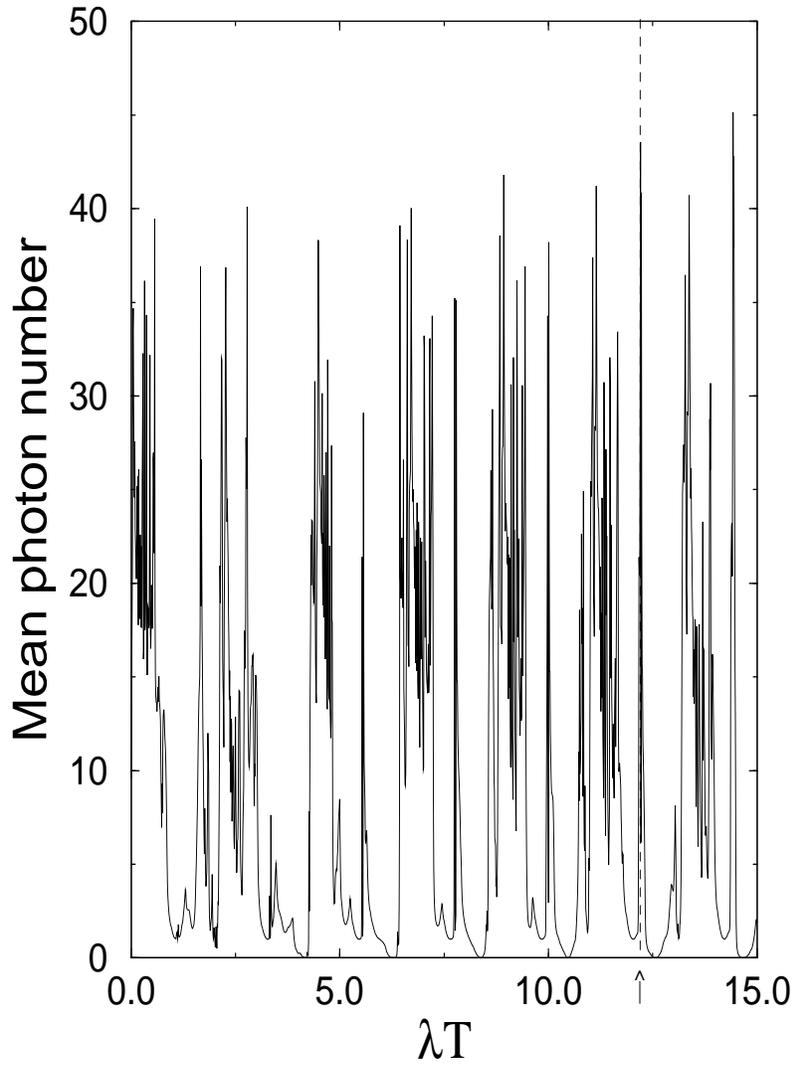,height=12cm,width=8cm}}
\vspace{2cm}
\caption{Mean photon number of the steady state field, as a function of the 
interaction times. The dashed line indicates $T=12.2/\lambda$.
The same parameters as in figure 1.}
\end{figure}

\newpage

\begin{figure}[hp]
\vspace{1cm}
\centerline{\hspace{1.0cm}\psfig{figure=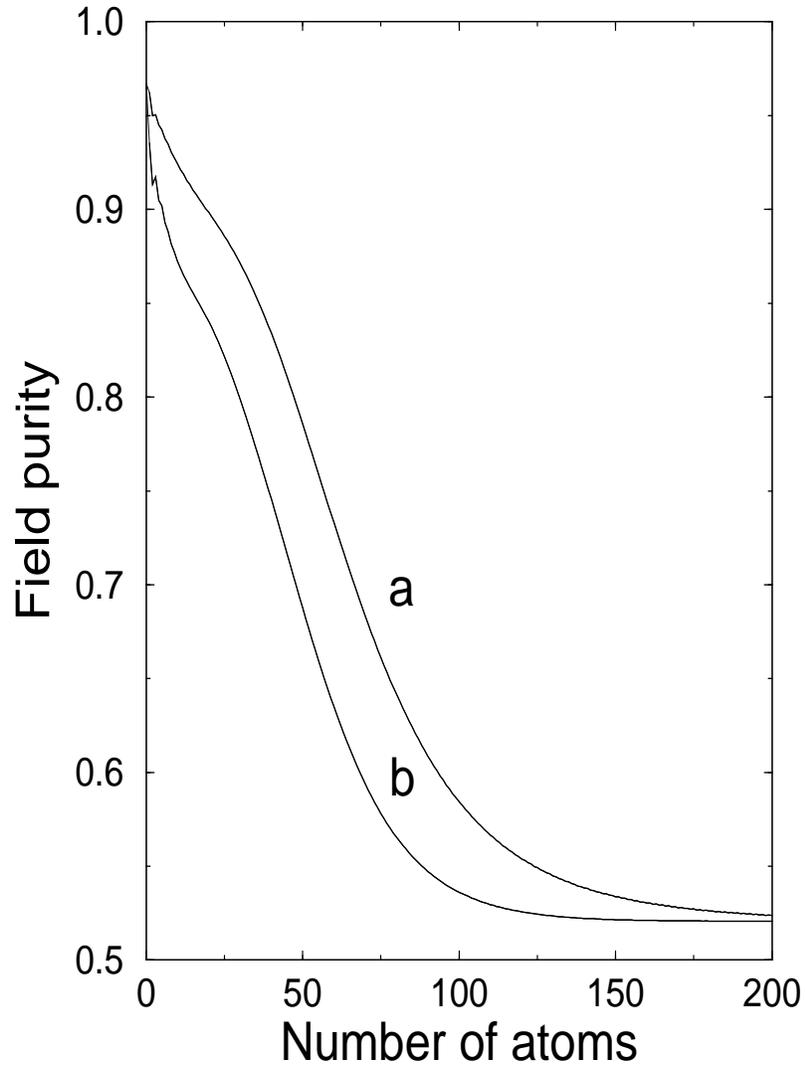,height=12cm,width=8cm}}
\vspace{2cm}
\caption{Purity of the field as a function of the number of atoms crossing the
cavity. (a) For an initial thermal state; (b) for an inital mixed coherent state.
Initial mean photon number $\overline{n}=10$.}
\end{figure}

\newpage

\begin{figure}[hp]
\vspace{1cm}
\centerline{\hspace{1.0cm}\psfig{figure=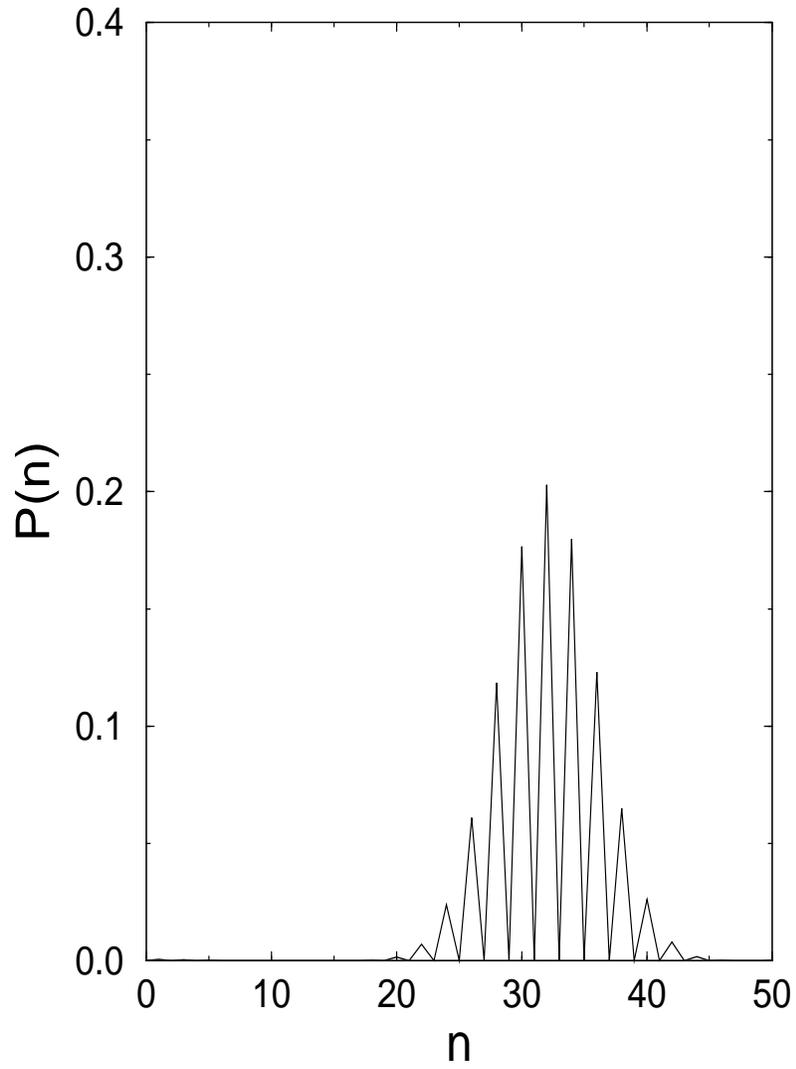,height=12cm,width=8cm}}
\vspace{2cm}
\caption{Photon number distribution of the steafy-state cavity field after 
having passed $N=200$ atoms, for an initial thermal field having $\overline{n}=10$.}
\end{figure}

\newpage

\begin{figure}[hp]
\vspace{1cm}
\centerline{\hspace{1.0cm}\psfig{figure=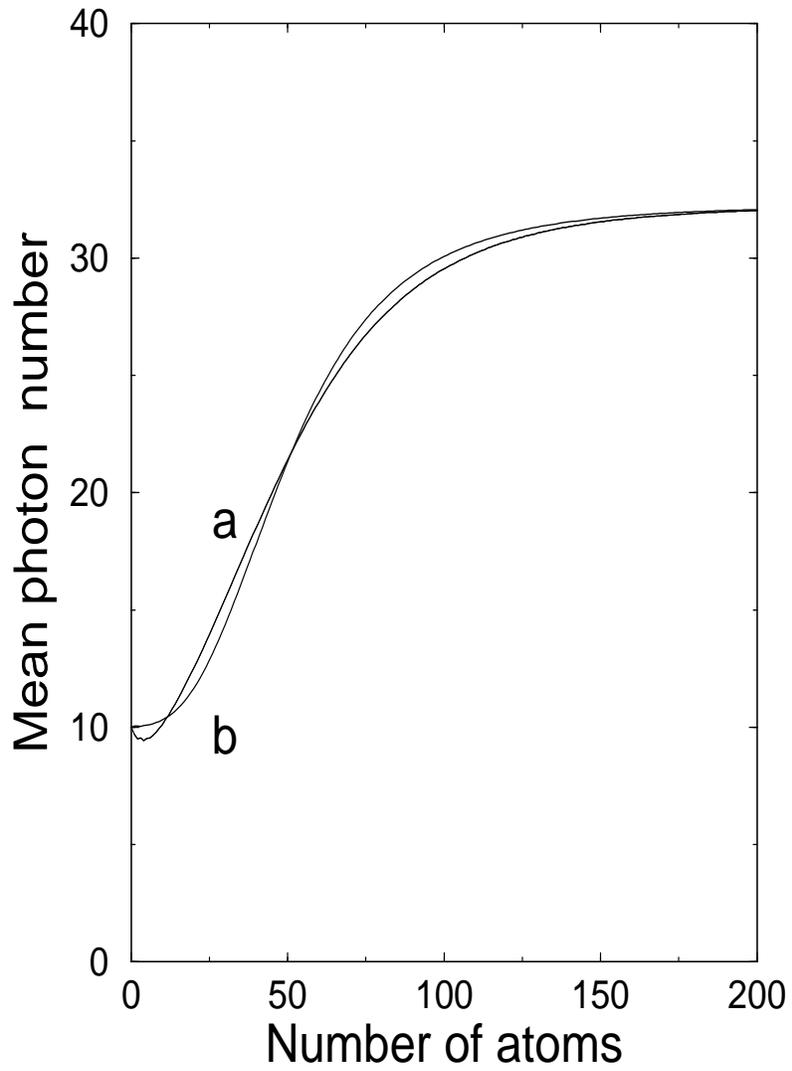,height=12cm,width=8cm}}
\vspace{2cm}
\caption{Mean photon number of the field as a function of the number of atoms
crossing the cavity. (a) For an initial thermal state; (b) for an inital 
mixed coherent state.}
\end{figure}

\newpage

\begin{figure}[hp]
\vspace{1cm}
\centerline{\hspace{1.0cm}\psfig{figure=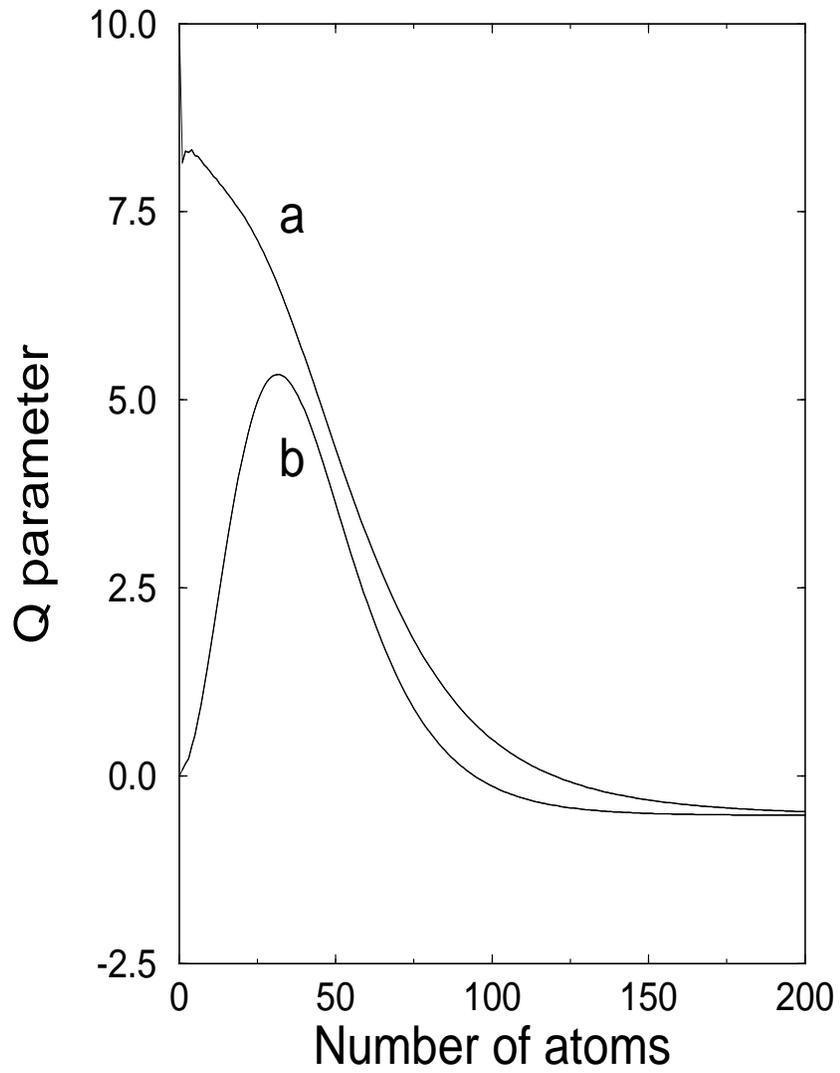,height=12cm,width=8cm}}
\vspace{2cm}
\caption{Mandel's Q parameter of the field as a function of the number of atoms
crossing the cavity. (a) For an initial thermal state; 
(b) for an inital mixed coherent state.}
\end{figure}

\newpage

\begin{figure}[hp]
\vspace{1cm}
\centerline{\hspace{1.0cm}\psfig{figure=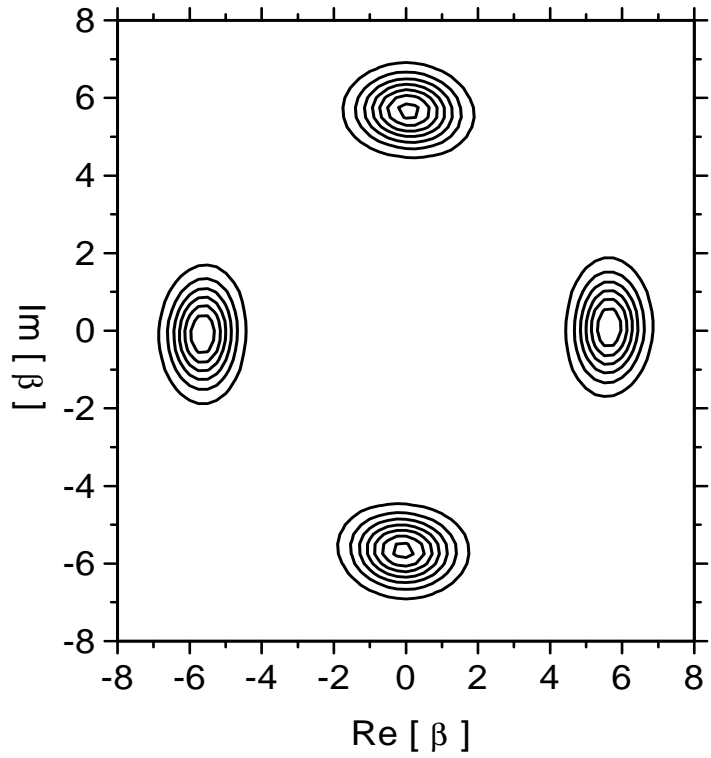,height=10cm,width=15cm}}
\vspace{2cm}
\caption{Q function contours of the cavity field after having passed $N=200$ atoms,
for an initial thermal field with $\overline{n}=10$.}
\end{figure}

\end{document}